\documentclass[twocolumn]{aastex631}

\shorttitle{Radio Counterpart to an Orion JuMBO}
\shortauthors{Rodr\'\i guez et al.}

\graphicspath{{./}{figures/}}

\begin{document}

\title{A Radio Counterpart to a Jupiter-Mass Binary Object in Orion}

\author[0000-0003-2737-5681]{Luis F. Rodr\'\i guez}
\affiliation{Instituto de Radioastronom\'\i{}a y
Astrof\'\i{}sica, Universidad Nacional Aut\'onoma de M\'exico \\
Apartado Postal 3--72, 58090 Morelia, Michoac\'an, M\'exico}
\affiliation{Mesoamerican Center for Theoretical Physics \\
Universidad Aut\'onoma de Chiapas \\
Carretera Emiliano Zapata Km 4, 29050 Tuxtla Guti\'errez, Chiapas, M\'exico}

\author[0000-0002-5635-3345]{Laurent Loinard}
\affiliation{Instituto de Radioastronom\'\i{}a y
Astrof\'\i{}sica, Universidad Nacional Aut\'onoma de M\'exico \\
Apartado Postal 3--72, 58090 Morelia, Michoac\'an, M\'exico}

\author[0000-0001-6553-4853]{Luis A. Zapata}
\affiliation{Instituto de Radioastronom\'\i{}a y
Astrof\'\i{}sica, Universidad Nacional Aut\'onoma de M\'exico \\
Apartado Postal 3--72, 58090 Morelia, Michoac\'an, M\'exico}

\begin{abstract}
Using James Webb Space Telescope near-infrared data of the inner Orion Nebula, \citet{Pearson_McCaughrean_2023} detected 40 Jupiter-Mass Binary Objects (JuMBOS). These systems are not associated with stars and their components have masses of giant Jupiter-like planets and separations in the plane of the sky of order $\sim$100 au. The existence of these wide free-floating planetary mass binaries was unexpected in our current theories of star and planet formation. Here we report the  radio continuum (6.1 and 10.0 GHz) Karl G.\ Jansky Very Large Array detection of a counterpart to JuMBO\,24. The radio emission appears to be steady at a level of $\sim$50 $\mu$Jy over timescales of days and years. We set an upper limit of $\simeq15$~km~s$^{-1}$ to the velocity of the radio source in the plane of the sky. As in the near-infrared, the radio emission seems to be coming from both components of the binary.
\end{abstract}


\section{Introduction} \label{sec:intro}

In a new James Webb Space Telescope (JWST) near-infrared survey of the inner Orion Nebula and Trapezium Cluster, \citet{Pearson_McCaughrean_2023} detected 40 Jupiter- Mass Binary Objects (JuMBOS).  Additionally, two triple objects were reported. These multiple systems are not associated to stars and their components have masses in the range of 0.6 to 14 Jupiter masses and separations in the plane of the sky between 28 and 384 au (0.07 and 0.99 arcsec at a distance of 388 pc; \citealt{Kounkel+2017}). The existence of these wide planetary-mass binaries is surprising since our current knowledge of star and planet formation cannot account for them. A possible theoretical explanation has been offered by \citet{Wang+2023} in terms of an ejection after a close stellar flyby between two solar-like stars. Rogue planets (single planets not gravitationally bound to any star or brown dwarf, also known as free-floating or orphan planets) have been detected in the optical and near-infrared \citep{Lucas_Roche_2000,ZapateroOsorio+2000}. This latter class of single sources, however, is compatible with the known physical conditions of forming stellar and planetary systems \citep{Hurley_Shara_2002,Kroupa_Bouvier_2003,Whitworth_Zinnecker_2004,MiretRoig2023}. We have searched for radio counterparts to the 42 JuMBOS in Orion using sensitive observations from the archive of the Karl G. Jansky Very Large Array (VLA) of the National Radio Astronomy Observatory. Here we report the radio detection of JuMBO\,24 in three epochs distributed across a decade, in 2012, 2018, and 2022. This paper is organized as follows. In Section 2 we present the observations, while in Section 3 we discuss the implications of this detection. Finally, in Section 4 we summarize our conclusions.

\section{Observations} \label{sec:obs}

\subsection{VLA Observations}

We analyzed the VLA data of the three projects listed in Table \ref{tab:obsvla}. The data were calibrated in the standard manner using the CASA (Common Astronomy Software Applications; \citealt{McMullin+2007}) package of NRAO and the pipeline provided for VLA\footnote{https://science.nrao.edu/facilities/vla/data-processing/pipeline} observations. A robust weighting of 0 \citep{Briggs_1995} was used in the imaging to optimize the compromise between sensitivity and angular resolution. The images were made with a $(u,v)$ range set to $>100~k\lambda$, to suppress structures larger than 2$''$ that limit the signal-to-noise ratio in the Orion field. The project SD0630 consists of five sessions made on 2012 September 30 and October 2, 3, 4, and 5. The projects 17A-069 and 22A-285 were made on single sessions on 2018 May 11 and 2022 July 2, respectively. In Table \ref{tab:obsvla} we also give the mean epoch of the observations, the configurations used, the central frequency and the bandwidth of the observations, the parameters of the synthesized beam, the radio flux density and position of JuMBO\,24, and the 4-$\sigma$ upper limit to the absolute circular polarization. The gain calibrator, J0541$-$0541, was the same for all observations so the astrometry can be directly compared more easily. The absolute flux calibrator was J0137$+$331 (3C 48) for projects SD0630 and 22A-285, and J0521$+$1638 
(3C 138) for project 17A-069.

We find that the near-infrared source JuMBO\,24 coincides within 0.05 arcsec with a radio continuum source present in all three VLA images.  The position of JuMBO\,24 given by \citet{Pearson_McCaughrean_2023} is RA(J2000) = $05^h 35^m 19\rlap.^s50616$, DEC(J2000) = $-05^\circ 23' 39\rlap.{''}7303$. The positions of the radio counterpart source for the three epochs are given in Table 1. This radio source first appeared in the list of \citet[][their source 441]{Forbrich+2016}. JuMBO\,24 was originally detected in near-infrared Hubble Space Telescope observations in the F160W (1.4-1.8 $\mu$m) and F110W (0.8-1.4 $\mu$m) filters, with magnitudes of 17.11 and 18.33, respectively \citep[][their source 274]{Luhman+2000}. \citet{Slesnick+2004} reported a K-band magnitude of 19.51 (their source 728). These authors used near-infrared photometry and spectroscopy as well as evolutionary stellar models \citep{Dantona_Mazzitelli_1994,Baraffe+1998} to determine an infrared spectral type of M5.5$\pm$1.5 and a mass of  $\sim 52$~M$_J$, under the assumption of a single star. The JWST data and the interpretation of \citet{Pearson_McCaughrean_2023} indicate that JuMBO\,24 is a binary with both components having the same giant-planet mass of $11.5~M_J$ and a separation of 28 au in the plane of the sky. JuMBO\,24 stands out in the list of these objects because it has the largest total mass ($23~M_J$) and the 
closest separation in the plane of the sky (28 au). Also, the primary of JuMBO\,24 has the second smallest extinction, $A_V$ = 3.6, behind only JuMBO\,27, that has $A_V$ = 2.4.

\begin{deluxetable*}{cccccccccc}
\tablenum{1}
\tablecaption{Parameters of the VLA Observations\label{chartable}}
\tablewidth{900pt}
\tabletypesize{\scriptsize}
\tablehead{
\colhead{} & \colhead{Mean} & 
\colhead{} & \colhead{$\nu$} & 
\colhead{$\Delta \nu$} & \colhead{Synthesized Beam} & 
\multicolumn{4}{c}{Observational results for JuMBO\,24}  \\ 
\colhead{Project} & \colhead{Epoch} & \colhead{Configuration} & \colhead{(GHz)} & 
\colhead{(GHz)} &
\colhead{($\theta_{max} \times \theta_{min}$; PA)} & \colhead{$F_\nu~(\mu$Jy)} & \colhead{RA(J2000)}  & \colhead{DEC(J2000)} &  $\vert V \vert$} 
\decimalcolnumbers
\startdata
SD0630 & 2012.762 & A, BnA$\rightarrow$A & 6.1 & 2.0 & $0\rlap.{''}25 \times 0\rlap.{''}21;+2^\circ$ & 45.4$\pm$4.5 & $19\rlap.^s5061 \pm 0\rlap.^s0006$ & 
$39\rlap.{''}730 \pm 0\rlap.{''}012$ &  $\leq 15\%$ \\
17A-069 & 2018.359 & A & 10.0 & 4.0 & $0\rlap.{''}21 \times 0\rlap.{''}17;-17^\circ$ & 60.0$\pm$13.0 
&  $19\rlap.^s 5065\pm0\rlap.^s0017$ & $39\rlap.{''}727\pm0\rlap.{''}024$ & $\leq 40\%$ \\
22A-285 & 2022.501 & A & 6.1 & 4.0 & $0\rlap.{''}27 \times 0\rlap.{''}21;+10^\circ$ & 50.0$\pm$7.2 &$19\rlap.^s 5059\pm 0\rlap.^s 0016$
&  $39\rlap.{''}698\pm0\rlap.{''}017$  & $\leq 20\%$ \\
\enddata
\tablecomments{Columns 4 and 5 are the central frequency and bandwidth of the observations. The position given in columns 8 and 9 show only the seconds of the RA and the arcseconds of the DEC, measured from RA(J2000) = $05^h35^m$ and DEC(J2000) = $-05^\circ23'$, respectively. Column 10 gives the 4-$\sigma$ upper limit to the absolute 
circular polarization.}
\label{tab:obsvla}
\end{deluxetable*}

\subsection{ALMA Observations}
We searched for JuMBO counterparts in the Atacama Large Millimeter/Submillimeter Array archive. 
We analyzed several data sets: 2013.1.00546.S, 2015.1.00534.S, 2015.1.00262.S, 2015.1.00669.S, 2017.1.01353.S,
2017.1.01639.S, and 2018.1.01107.S. However, we do not find any evident millimeter or even submillimeter
continuum counterpart for any of the 42 JuMBOs. In particular, the data set 2015.1.00669.S 
(PI: Hacar, Alvaro) is a large mosaic made in Band 3 that covers a large area 
(OMC2, OMC1-N, OMC1, OMC1-S).
The rms noise for this large mosaic is 150 $\mu$Jy Beam$^{-1}$, 
so we do not detect any counterpart at
a 5-$\sigma$ level of 750 $\mu$Jy Beam$^{-1}$. Future much deeper ALMA observations of the Orion Nebula could
reveal thermal emission related with the JuMBOS.

\section{Discussion} \label{sec:discussion}

The first centimeter detection of a single planetary mass object, SIMP J01365662+0933473, was achieved 
by \citet{2018ApJS..237...25K}.
The first identification of a centimeter continuum source with a planetary-mass binary object presented
here is also of great relevance by the information the radio data can provide. We exemplify these possibilities in what follows. Further work, particularly in relation to the origin of the radio emission is deferred to forthcoming papers.

\subsection{The Probability of a Random Association}

In this section we estimate how likely it is that the positional agreement of JuMBO 24 with one of the radio sources in the region is fortuitous and not real. In their study of Orion, \citet{Forbrich+2016} detected a total of 556 radio sources in a region of $13\rlap.’6 \times 13\rlap.’6$. Most of the sources, however, are enclosed in a region with radius of $\sim 3’$
(see Figure 5 of \cite{Forbrich+2016}). Of the 556 sources, 358 have flux densities equal or larger than 50 $\mu$Jy, the characteristic value for the radio source proposed to be associated
with JuMBO 24. These results give a solid angle density of $3.5 \times 10^{-3}$ radio sources per arcsec$^2$.

The most accurate position for the radio source is that obtained from the multiepoch project SD0630. Convolving the statistical errors of the position of the radio source (Table 1) with the systematic positional error of $0\rlap.{''}01$, as estimated by \citet{Dzib+2017}, we obtain a 
$\pm 1-\sigma$ error box of
$0\rlap.{''}027 \times 0\rlap.{''}031$. The \sl a priori \rm probability of a source randomly falling in this error box is then $2.9 \times 10^{-6}$. However, the coincidence of the radio source with any of the 42 JuMBOs in Orion would have been considered significant, so assuming that these possible coincidences are independent events we have to multiply the probability of a given JuMBO source being spatially associated times 42. We finally get an \sl a priori \rm probability of one JuMBO randomly coinciding in position with one radio source to be only $1.2 \times 10^{-4}$, supporting a real association.

\subsection{Upper Limits to the Proper Motion}

Using the radio positions determined at the three epochs we can search for the proper motions of the source. As noted before, the fact that the same nearby ($\sim 1\rlap.^\circ5$) gain calibrator (J0541$-$0541) was used in all observations implies a sound astrometry. A least-squares fit to the
positions given in Table 1, after adding in quadrature a systematic error of $0\rlap.{''}01$, yields the following values:

$$\mu_{\alpha} \cos \delta = +1.8 \pm 2.8 \text{~mas~yr}^{-1} ~~; ~~\mu_\delta  = -3.0 \pm 2.5\text{~mas~yr}^{-1}.$$

\noindent
This result implies that there are no
significant proper motions at a 3-$\sigma$ level of $\sim$8 mas~yr$^{-1}$, that corresponds to a transverse velocity of $\sim 15$ km~s$^{-1}$ at a distance of 388 pc \citep{Kounkel+2017}. These upper limits rule out large velocities in the plane of the sky, although it should be pointed out that significant velocities could be
present in the radial direction. Our measurements can be registered to an approximate rest frame for the Orion Cluster by subtracting the average proper motion of the radio stars in the core of Orion \citep{Dzib+2017},
$\mu_{\alpha} \cos\delta = +1.07 \pm 0.09\text{~mas~yr}^{-1}; \mu_{\delta}  = -0.84 \pm 0.16\text{~mas~yr}^{-1}$. We obtain:

$$\mu_{\alpha} \cos \delta = +0.7 \pm 2.8\text{~mas~yr}^{-1} ~~ ; ~~ \mu_\delta  = -2.2 \pm 2.5\text{~mas~yr}^{-1}.$$

\noindent
We conclude that JuMBO\,24 is not moving at large velocities compared to the radio stars in Orion. \citet{2019A&A...624A.120V} have made numerical simulations for the evolution of young and dense stellar clusters.
They find that single free-floating planets ejected from their original system after a strong encounter with another star will have velocities typically three times those of the stars in the cluster. \citet{Dzib+2017} find a typical 
velocity dispersion of $\sim 2-3$ km~s$^{-1}$ along a plane-of-the-sky coordinate
for the radio-emitting stars in the core of the Orion Nebula Cluster.
Assuming that the results of \citet{2019A&A...624A.120V} can be extended to free-floating binary planets, we expect
velocity dispersions of $\sim 6-9$ km~s$^{-1}$ for the JuMBOs as well as for the $\sim$500
single planetary-mass objects
reported in Orion by \citet{Pearson_McCaughrean_2023}.
Future highly accurate determinations of the kinematics of the Orion JuMBOs will help understand the origin of these
unexpected sources.


\subsection{Time Monitoring}

The two measurements at 6.1 GHz given in Table 1 are separated by a decade and, within the noise, they are consistent with being equal. Another evidence that weighs in favor of a steady flux is the analysis of the five different epochs observed in project SD0630. In Table 2, we give the epochs and flux densities for each individual observing session, where we can see that all flux densities are consistent within noise with a value of $\simeq 45~ \mu$Jy. While more intensive observations are needed to establish firmly the time behavior of the radio emission from JuMBO\,24, all existing data are consistent with a steady flux density over timescales of days to years.

\begin{deluxetable}{ccc}
\tablenum{2}
\tablecaption{Flux Densities of JuMBO\,24 at 6.1 GHz from Project SD0630\label{tab:time}}
\tablewidth{400pt}
\tabletypesize{\small}
\tablehead{
\colhead{} & \colhead{Mean UTC} & \colhead{Flux Density} \\
\colhead{Date} & \colhead{(hours)} & \colhead{($\mu$Jy)}}
\decimalcolnumbers
\startdata
2012 Sep 30 & 11.3 & 52.9$\pm$6.5   \\
2012 Oct 02 & 11.2 &   39.7$\pm$5.5 \\
2012 Oct 03 & 14.8 &  55.0$\pm$11.0  \\
2012 Oct 04 &  13.4 &  44.9$\pm$7.1  \\
2012 Oct 05 & 12.3  &  40.5$\pm$6.6 \\
\enddata
\end{deluxetable}

\subsection{Nature of the Radio Emission} \label{spinx}

The objects with known radio emission that are more similar to the components of the JuMBOs are the ultracool dwarfs \citep{2022ApJ...932...21K}. The radio emission from the latter objects has two main components: a periodically bursting emission arising in aurorae and a slowly varying, quiescent emission with a low
degree of circular polarization arising from a structure
similar in morphology to the Jovian radiation belts \citep{2023Sci...381.1120C, 2023Natur.619..272K}.
The bursting component is circularly polarized and it is believed to be produced by the
electron cyclotron maser instability \citep{2006ApJ...653..690H}, 
the same coherent emission process that creates the auroral radio emission from the planets of the Solar System 
\citep{2017ApJ...846...75P}. 
On the other hand, the quiescent component is usually interpreted as being caused by radio emission from relativistic electrons giving rise to gyrosynchrotron or synchrotron mechanisms, according to how relativistic the electrons are.
Now it is known that this quiescent emission is produced in radiation belts \citep{2023Sci...381.1120C, 2023Natur.619..272K}. Given that the radio emission from JuMBO 24 is steady in time, one would expect it to be 
produced by relativistic electrons spiraling in the magnetic fields of a radiation belt.

The spectral index of the radio emission is an indicator of the nature of the radiation. The observations of SD0630 were made using two bandwidths of 1~GHz each, centered at 4.8 and 7.4 GHz, respectively. We obtain that the flux densities at these two frequencies are 53.8$\pm$6.0 and 43.0$\pm$5.8 $\mu$Jy, respectively. Assuming that the continuum spectrum is described by a power law, $S_{\nu} \propto \nu^{\alpha}$, we derive a spectral index of $\alpha = -0.52\pm0.40$. We can also derive a spectral index assuming that the source has a steady flux density and combining the values given in Table 1. From these data we derive $\alpha = +0.50\pm0.43$. The weighted mean and rms of these two
determinations is $\alpha = -0.05\pm0.51$.
Given the large uncertainty, it is difficult to discriminate between possible emission mechanisms, but we can rule out an optically-thick blackbody in the Rayleigh-Jeans regime ($\alpha$ = 2) and the negative steep indices characteristic of pulsars ($\alpha \leq$ --2). 

We did not detect circular polarization at any epoch. However, the upper limits are not very stringent. The presence of circular polarization is a strong indicator of either electron cyclotron maser instability,
cyclotron or gyrosynchrotron radiation but, of course, its absence does not rule out these mechanisms.

As noted above, the ultracool dwarfs are the closest thing to giant planets that have been detected in the radio. 
The radio luminosities of nine quiescent binary ultracool dwarf systems discussed by
\citet{2022ApJ...932...21K} are in the range of $10^{11.9-14.6}$ erg s$^{-1}$
Hz$^{-1}$. JuMBO 24 has a radio luminosity of $10^{16.0}$ erg s$^{-1}$
Hz$^{-1}$. We then conclude that the radio luminosity of JuMBO 24 is orders of magnitude larger than naively expected.

\subsection{Position and Morphology of the Radio and Infrared Sources}

\begin{figure*}
\plotone{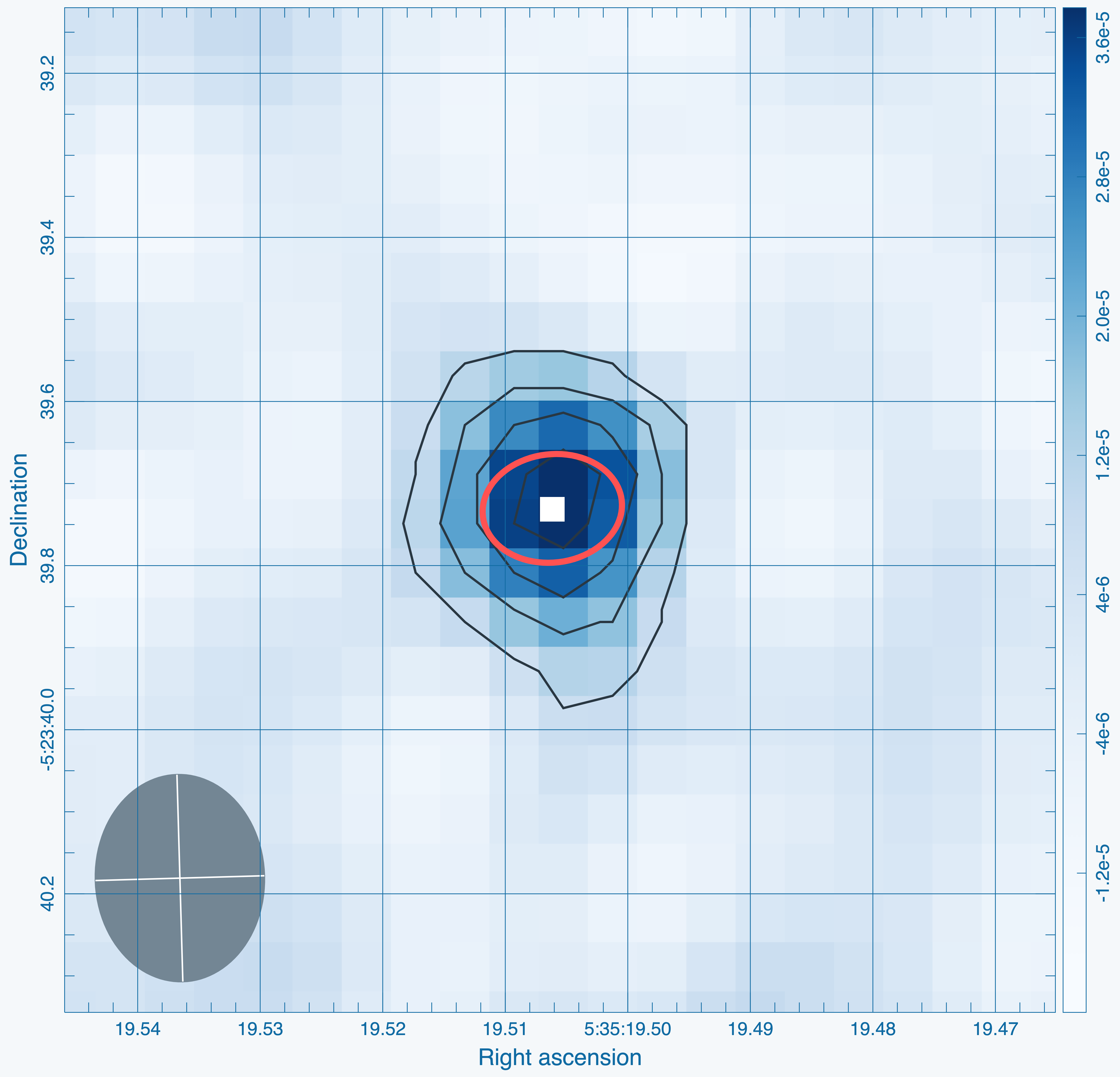}
\caption{VLA image of the 6.1 GHz emission from JuMBO\,24, made concatenating all the data of project SD0630. The contours start at, and increase by steps of, 9 $\mu$Jy~beam$^{-1}$. The color scale indicates intensity as shown by the color bar to the right. The white rectangle indicates the position of JuMBO\,24 as reported by \citet{Pearson_McCaughrean_2023}: RA(J2000) = $05^h 35^m 19\rlap.^s50616$, DEC(J2000) = $-05^\circ 23' 39\rlap.{''}7303$. The red ellipse shows the angular size and orientation
of the target in the 2.77 $\mu$m JWST images ($0\rlap.{''}170 \times 0\rlap.{''}132; 97^\circ$). The synthesized beam of the radio map is shown at the bottom left of the image.}
\label{fig:VLAmap}
\end{figure*}

In Figure \ref{fig:VLAmap} we show an image of the radio continuum emission obtained from concatenating all the data of project SD0630. We can see that the radio source coincides closely with the near-infrared position given by \citet{Pearson_McCaughrean_2023}. A Gaussian ellipsoid fit to the radio image using the task IMFIT of CASA gives a deconvolved angular size of $0\rlap.{''}179 \pm 0\rlap.{''}045 \times 0\rlap.{''}143  \pm 0\rlap.{''}074; 60^\circ \pm 89^\circ$. Although the radio source is only marginally resolved, these angular
dimensions and the orientation of the major axis are consistent with the value obtained from Gaussian ellipsoid fits to the JWST images. For instance a fit to the image at 2.77 $\mu$m results in $0\rlap.{''}170 \pm 0\rlap.{''}002 \times 0\rlap.{''}132  \pm 0\rlap.{''}002; 97^\circ \pm 2^\circ$. The shorter wavelength JWST observations show that the cause of the extension along the east-west direction is the binarity of the source, which is comprised of two objects separated by about 100 mas precisely along the east-west direction. This result suggests that, as in the near-infrared, there is significant radio emission from both components of the binary. This result is similar to that obtained for the compact ultra-cool dwarf binary VHS 1256$-$1257AB, where the optical and radio emissions show compatible dimensions and orientation \citep{Rodriguez+2023}.

\section{Conclusions}

We report the detection of radio continuum emission at 6 to 10 GHz associated with the recently identified Jupiter-Mass Binary Object JuMBO\,24 in the Orion Nebula Cluster. Within the errors, the radio emission appears steady on timescales from days to years and exhibits no circular polarization. No significant proper motion is detected between the three observing epochs, ruling out large velocities ($\geq 15~km~s^{-1}$) in the plane of the sky. The radio emission is marginally resolved in the same direction as the infrared source detected by the JWST, suggesting that the radio emission comes from a combination of the two planetary mass objects. Additional radio observations are necessary to pin down the nature of the radio emission mechanism.

\bibliography{JuMBO24}{}
\bibliographystyle{aasjournal}

\begin{acknowledgments}
We thank the anonymous reviewer for several valuable comments that improved the paper.
The National Radio Astronomy Observatory is a facility of the National Science Foundation operated under cooperative agreement by Associated Universities, Inc. We acknowledge the support from DGAPA, UNAM (projects IN112323, IN112417, IN112820 and IN108324) and CONAHCyT, México (projects 238631, 280775 and 263356).
\end{acknowledgments}

\vspace{5mm}
\facilities{NASA/ESA/CSA(JWST), NRAO(VLA)}

\software{AIPS \citep{Greisen_2003}, CASA \citep{McMullin+2007}}

\end{document}